\documentclass[twocolumn,showpacs,preprintnumbers,amsmath,amssymb]{revtex4}
\usepackage{graphicx}% Include figure files
\usepackage{dcolumn}% Align table columns on decimal point
\usepackage{bm}% bold math
\DeclareGraphicsExtensions{.pdf,.png,.jpg}
\usepackage{epstopdf}
\begin{document}

\title{Intra-inter band pairing, order parameter symmetry in Fe-based superconductors : A model study}
\author{Smritijit Sen, Haranath Ghosh\footnote{Both the authors have equal contribution in this work.}}
\affiliation{Indus Synchroton Utilization Division, 
Raja Rammana Centre for Advanced Technology, Indore, 452013, India.}

%\author{}

%\address{}

\begin{abstract}
In the quest of why there should be a single transition temperature in a multi-gapped
system like Fe-based materials we use two band model for simplicity. The model comprises 
of spin density wave (SDW), orbital density wave (ODW) arising due to nested pieces of the
 electron and hole like Fermi surfaces;
 together with superconductivity of different pairing symmetries around electron and hole 
like Fermi surfaces. We show that either only intra or only inter band pairing is insufficient 
to describe some of the experimental results like large to small gap ratio, thermal behaviour 
of electronic specific heat jump etc. It is shown that the inter-band pairing is essential in 
Fe-based materials having multiple gaps to produce a single global $T_c$. Some of our results 
in this scenario, matches with the earlier published work \cite{two-band-prb}, 
and also have differences. 
The origin of difference between the two is also discussed. 
Combined intra-inter band pairing mechanism 
produces the specific heat jump to superconducting transition temperature ratio proportional 
to square of the transition temperature, both in the electron and hole doped regime, for sign 
changing s$^{\pm}$ wave symmetry which takes the d+s pairing symmetry form. Our work thus 
demonstrates the importance of combined intra-inter band pairing irrespective of the pairing 
mechanism.
\end{abstract}

%\begin{keyword}
%Fe-based superconductor, Two band model, Specific heat, Characteristic ratio, pairing symmetry.
%%\MSC[2008] code \sep code (2000 is the default)

%\end{keyword}

%\end{frontmatter}

\maketitle
% \linenumbers
\section{Introduction}
\label{sec1}

Recent discovery of high temperature superconductivity at 26 K in LaFeAsO doped 
with F on the oxygen site in 2008 is of immense importance \cite{Kamihara} in the history of
superconductivity. These new types of superconductors
have conducting layers of iron and a pnictide (Pn)/chalcogenide (Ch)
(typically arsenic/selenium) and seems to show great potential as the next 
generation high temperature superconductors.
Dominance of Fe electrons at the Fermi surface (FS) and unusual Fermiology, that can be 
modulated by doping, makes normal and SC state properties of 
iron-based superconductors quite unique compared to those of conventional 
electron-phonon coupled superconductors \cite{Stewart}. These Cu (sometimes also O) free new 
compounds are different from the high $T_c$ cuprates and may lead to a non-BCS 
(Bardeen, Cooper and Schrieffer Theory) type superconductivity with a better 
theoretical and experimental understanding on the mechanism of unconventional 
high-$T_c$ superconductivity. Importance of mutual influences of electronic spin degrees 
of freedom (magnetism), orbital degrees of freedom (orbital order) and pairing symmetry
in superconductivity
can not be overemphasised, all these play a special role in Fe-based materials \cite{Chen,Nandi S}.
Pairing mechanism and information about the pairing symmetry of the cooper pair 
wave-functions are the key ingredients for developing a theory of these iron-based 
superconductors. The total electronic wave function of the cooper pairs must be 
antisymmetric under their exchanges. Therefore, for spin singlet state ($S=0$) which
is antisymmetric, its orbital wave function would be symmetric, leading to
 s-wave, d-wave, g-wave type orbital natures.
In contrast, for spin triplet state ($S=1$), its spin wave function being symmetric, 
its orbital wave function would be anti-symmetric (p-wave, f wave etc.). 
In the conventional low $T_c$ superconductors ({\it e.g}., Pb, Al, Hg, Nb, Nb$_3$Sn etc.),
 the phonon mediated electron-electron interaction leads to spin singlet pairing 
with s-wave symmetry. On the other hand, the pairing symmetry of cooper pairs in the 
high T$_c$ cuprate superconductors is dominantly d$_{x^2- y^2}$ kind and it corresponds to 
$l$=2 orbital angular momentum \cite{Tsuei,Van Harlingen,hng1,hng2,hng3}. With significantly improved sophisticated 
experimental and theoretical tools, the question of pairing symmetry in Fe-based 
superconductors is thoroughly studied and there are enough experimental evidences 
for some version of the so-called s$^\pm$ state \cite{Mazin,Barzykin,Chubukov}, although predictions of other 
pairing states like s$^{++}$ state mediated by orbital fluctuations  are also available 
in the literature \cite{Kontani,Yanagi}.
 However, order parameter (OP) symmetry and the pairing mechanism are far from being settled. 
Neutron scattering experiments provide convincing indication for a sign changing 
SC energy gap $\Delta (k)$ on different parts 
of the FS in a number of iron based superconductors \cite{chris}.
Experimental studies on the SC gap in iron-based superconductors reveal that 
there are two nearly isotropic gaps with characteristics ratios 
2$\Delta_{SC}$(k)/k$_B$T$_c$ = 2.5 $\pm$ 1.5 (for small gap on the 
outer $\Gamma$-barrel) and 7$\pm$2 (on the inner $\Gamma$-barrel and the propeller-like structure around the X point, for large gap) which is considerably 
different from the conventional BCS characteristic ratio 3.5 \cite{Evtushinsky}.
The behaviour of specific heat of these iron-based superconductors is also distinctly 
different. For conventional BCS superconductors, the electronic specific heat (C$_e$)
 decreases 
exponentially with decrease of temperature below T$_c$. But in case of iron-based 
superconductors the electronic specific heat decreases with decreasing temperature 
below $T_c$ obeying power law. In general, specific heat data not only reveals the 
SC transition at lower temperatures but also about the higher temperature 
transitions, like structural and magnetic 
[for example, spin density wave (SDW), orbital density wave (ODW)] transitions.
If enough magnetic field is applied to conquer T$_c$ appreciably, C/T extrapolated 
to $T=0$ from normal state data provides Sommerfeld constant 
$\gamma_n \equiv \lim_{T\to 0} ~ C_{normal}/T$, 
which is proportional to the renormalized bare electron density of states at the 
Fermi energy N(0); {\it i.e.}, $\gamma_n\sim(1+\lambda)$N(0), (where $\lambda$ can be a combination of 
electron-phonon and electron-electron interactions). It is a very useful parameter 
exploitable from specific heat data, as it is related to band structure calculations, 
resulting density of state N(0). Furthermore, the same is also related to the de 
Haas van Alphen measurement of effective masses of various FS orbits 
($\gamma_n\propto m^*)$. Because of large phononic contribution at higher temperatures, 
the specific heat jump ($\Delta$C) is not clear in some cases. 
If the phonon contribution to the specific heat below T$_c$  can be accurately estimated, 
e.g., via substitution of a neighbouring composition (replacing Fe by Co doping as they have almost 
same molar mass) that is not superconducting, one can extrapolate the 
electronic specific heat (C$_e$) below  T$_c$ and calculate $\gamma_n$. Another important 
parameter that correlates $\Delta$C and T$_c$ is $\Delta$C/T$_c$, and dependence of 
$\Delta$C/T$_c$ with T$_c$ for iron-based superconductors is again quite different from 
all other classes of superconductors including electron-phonon coupled conventional 
superconductors. Bud'ko, Ni and Canfield (BNC) plotted $\Delta$C/T$_c$
 as a function of $T_{c}^2$ for 14 different samples of various doped 
BaFe$_2$As$_2$ superconductors which indicate $\Delta C/T_c = aT_{c}^2$
 with  $a ~\sim$ 0.056 mJ/mole-K$^4$ \cite{Bud’ko}.   
Later on J. S. Kim \textit{et al}., modified BNC plot to include all other FePn/Ch 
superconductors and showed $\Delta C/T_c = aT_{c}^{1.9}$ with  $ a ~\sim$ 0.083 mJ/mole-K$^4$ 
\cite{Kim}   whereas the electron-phonon coupled conventional superconductors
 show significantly different temperature dependence (e.g., $\Delta C/T_c \propto T_c$). 
In this respect also Fe-based materials are unique, in the sense that none of the so far known 
earlier classes (like conventional BCS, A-15, heavy fermion, high T$_c$ cuprates etc.) 
of superconductors follow $\Delta C/T_c \propto T_{c}^2$. 
\par In this work, we use the minimal two band model ($d_{xz},d_{yz}$) of superconductivity 
in three different scenarios: (i) intra band pairing (ii) inter band pairing and (iii) 
combined intra-inter band pairing on equal footing to study Fe-based superconductors. 
In case of intra band pairing two distinctly different $T_cs$ are obtained which does not
 meet the experimental finding of single $T_c$ from angle resolved photo emission studies 
(ARPES). Therefore, only intra band pairing is not sufficient to describe Fe-based materials 
and hence excluded from our calculations. In the inter-band only pairing potential, 
single $T_c$ is obtained. In this picture, we present our analytical results of integral gap 
equations involving all the orders like SDW, ODW, and superconducting (SC) gaps around electron, 
hole Fermi surfaces. We show that in the limiting case of vanishing SDW, ODW orders, the SC gap 
equations reproduce similar form as published in \cite{two-band-prb}. Therefore, our work is 
more generalization of the work \cite{two-band-prb}  including SDW and ODW orders. We show that 
only inter-band pairing interaction of superconductivity can not produce specific heat jump 
such that $\Delta C/T_c \propto T_c^2$. Thus, as suggested in \cite{two-band-prb}  
we consider both intra-band and inter-band pairing on an equal footing which reproduces some 
of the experimental features like the ratio of large gap/small gap at T=0K (that inter-band 
picture fails to produce).
We show that the behaviour of $\Delta C/T_c$ with T$_c$ and the estimated values of 
2$\Delta_{SC}/k_BT_c$ are consistent with the experimental observations on 122 family 
of FePn in the  combined intra-inter band pairing picture. From our theoretically 
calculated data we found two jumps in the thermal 
variations of electronic specific heat, one at low temperature (SC transition) 
and another at higher temperature (SDW and ODW transition). We also calculate the value
 of 2$\Delta_{SC}/k_BT_c$ within two band model of Fe-based superconductors (both electron 
and hole doped situation), for all possible allowed pairing symmetry from the temperature 
dependent superconducting order parameters (SCOP). We further studied in detail, the behaviour of 
specific heat as a function of temperature for all possible allowed pairing symmetries 
like isotropic s-wave, d+s, s$_{xy}$ etc. In each case, we have calculated the value of 
$\Delta C/T_c$ as a function of T$_c$ which matches nicely with experimental 
behaviour. Through these model calculations we argue that, both inter as 
well as intra band pairing (irrespective of pairing mechanism) is required 
to explain some of the observed data. Rest of the paper is organized as 
follows. In the next section we describe our theoretical model describing 
its essential ingredients leading to the detailed calculations of the 
various OPs which are then used to calculate specific heat. 
In the results and discussion section we discuss our detailed results and 
finally conclude in the conclusion section.

\section{Theoretical Model}

First principle band structure calculations reveal that the density of states 
near
Fermi level dominantly have Fe-3d character. Among all these Fe orbitals, 
3d$_{yz}$,3d$_{xz}$ have the most contribution to the density of states at the 
Fermi level \cite{Tao}. Cao \textit{et al}., \cite{Cao} used 16 localized 
Wannier functions to 
build a tight binding effective Hamiltonian. Kuroki \textit{et al}., 
\cite{Kuroki} have used a five 
orbital tight binding model to explain the nature of band structure near the 
Fermi energy. S. Raghu \textit{et al}., \cite{Raghu} suggested a minimal 
two-band model that generates 
a topologically similar FS observed experimentally. We use two orbitals 
(d$_{xz}$,d$_{yz}$) per site on a two dimensional square lattice of iron. 
We take the mean field model Hamiltonian within the two band picture as 
\cite{Ghosh},  
\begin{eqnarray}
\label{ham}
H &=&\sum_{k,\sigma}^{FBZ}\varepsilon_{k}^e C_{k,\sigma}^\dagger C_{k,\sigma}
+
\sum_{k,\sigma}^{FBZ}\varepsilon_{k}^h f_{k,\sigma}^\dagger f_{k,\sigma} +
\Delta_{SDW}\sum_{k,\sigma}^{FBZ} \nonumber \\ &&
 (C_{k,\sigma}^\dagger 
\sigma_{\sigma,\sigma^\prime}^{z}
f_{k+Q,\sigma^\prime}+h.c.) 
-i\sum_{k,\sigma}^{FBZ}\Delta_{ODW} 
(C_{k,\sigma}^{\dagger}
f_{k+Q,\sigma} \nonumber \\ &&
- f_{k+Q,\sigma}^{\dagger}C_{k,\sigma})+
%+\sum_{k}\Delta_{SC}^e (k)( C_{-k,\downarrow}C_{k,\uparrow} %\nonumber \\
%+h.c.)
%+ \nonumber \\ && 
\sum_{k}^{FBZ}\Delta_{SC} (k)( C_{-k,\downarrow}C_{k,\uparrow}+ \nonumber \\ &&
f_{-k,\downarrow}f_{k,\uparrow}+h.c.)~~~~
\end{eqnarray}  
\begin{figure}[ht]
%\epsfxsize=4.0truein
%\epsfysize=3.5truein
%\epsffile{FS.eps}
\centering
\flushleft\includegraphics[width=8.5cm]{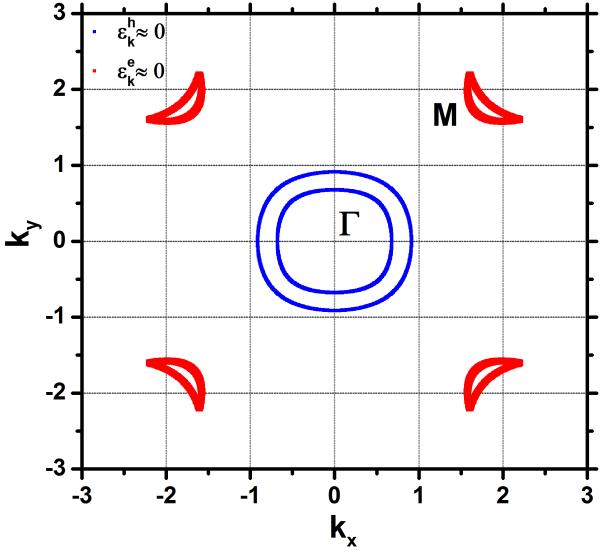}
\caption{Fermi Surface (or zero energy contour) in the reduced + rotated brillouin zone for 
$\mu$ =1.42 evaluated within two band model.}
\label{FS}
\end{figure}
The first two terms of the above Hamiltonian represent kinetic (band)  
energies in the electronic ($C_{k,\sigma}$ being the annihilation operator of an electron 
with spin $\sigma$) and hole ($f_{k,\sigma}$ being the annihilation operator of a hole) bands
around the four corners M and $\Gamma$ points respectively (see FIG.\ref{FS}). The electronic
and hole band dispersions are obtained as, \cite{Raghu,Ghosh,SSPS13} $\varepsilon_{k}^{e/h}= 
\epsilon_{+}(k)\pm \sqrt{\epsilon_{-}^2(k)+\epsilon_{xy}^2(k)} - \mu$ where 
$\epsilon_{+}(k)=-(t_1+t_2)(\cos k_x +\cos k_y) - 4 t_3 \cos k_x \cos k_y$
for two band model. 
The OPs $\Delta_{SDW}, \Delta_{ODW}$ represent respectively the spin 
density wave (SDW) and orbital density wave (ODW) 
that involves ordering between the electron and hole like bands (that are
nested by the nesting vector Q = (0, $\pi$) or ($\pi$,0). This ingredient in our model that the
 electron-like FS nests with the hole-like one and vice versa, is justified as it is
 consistent with recent experimental finding \cite{natureFS}. In ref \cite{natureFS} weak z-direction dispersion among the 
$\Gamma$ barrel and electron FSs are found resulting quasi-2d nested nature \cite{FS}.
For further details see below. The fifth term represent 
the terms involving superconductivity (SC) where $\Delta_{SC}\left(k\right)=(\Delta_{SC}^{e}\left(k\right)-
\Delta_{SC}^{h}\left(k\right))$; $\Delta_{SC}^{e/h}$ being SCOP around 
the electronic and hole FSs respectively. Our model consideration of 3d$_{yz}$,3d$_{xz}$ orbitals for superconductivity is also consistent with very recent finding of electron pairing at Fe-3d$_{yz,xz}$ orbitals \cite{PRB2014}.
The most general form of on-site interaction Hamiltonian for two band model may be obtained
as,
$H_{on-site}= \frac{1}{2}\sum_{i,\sigma,\sigma^\prime} 
\sum_{\alpha,\beta,\gamma,\delta=c,f} U_{\gamma,\delta}^{\alpha,\beta} 
\Psi_{i,\alpha,\sigma}^\dagger \Psi_{i,\beta,\sigma}^\dagger \Psi_{i,\delta,\sigma^\prime} \Psi_{i,\gamma,\sigma^\prime}$
where, $\Psi_{i,c,\sigma}^\dagger = C_{i,\sigma}^\dagger$ and $\Psi_{i,d,\sigma}^\dagger = f_{i,\sigma}^\dagger$
as used in the Hamiltonian (1) in momentum representation. Several intra and inter pocket 
electron-electron repulsion terms exists and according to the formulation \cite{Podolosky}, 
the mean field theory of SDW and ODW is obtained considering the mean field 
 OPs as, 
$\Delta_{SDW} = -U_{inter}\sum_{k,\sigma}^{FBZ}<C_{k,\sigma}^\dagger f_{k+Q,\sigma}+ h.c.>$ and 
$\Delta_{ODW} = -iV_{ODW}\sum_{k,\sigma}^{FBZ}<C_{k,\sigma}^\dagger f_{k+Q,\sigma} - 
f_{k+Q,\sigma}^\dagger C_{k,\sigma}>$ where both the $U_{inter}$ and $V_{ODW}$ are related 
to $U_{cc}^{dd}$ (see for details \cite{Podolosky}). 
Typical terms corresponding to superconductivity are given as follows, $H_{SC} = 
\sum_{k,k^\prime}^{FBZ} (V_{k,k^\prime}^eC_{k\uparrow}^\dagger C_{-k\downarrow}^\dagger 
C_{-k^\prime\downarrow} C_{k^\prime\uparrow}+V_{k,k^\prime}^h
f_{k\uparrow}^\dagger f_{-k\downarrow}^\dagger  
f_{-k^\prime\downarrow} f_{k^\prime\uparrow}+ 
V_{k,k^\prime}^{e-h}C_{k\uparrow}^\dagger C_{-k\downarrow}^\dagger f_{-k^\prime\downarrow} 
f_{k^\prime\uparrow} +h.c.)$. The first two terms correspond to intra-band pairing and the
pairing interaction $V_{k,k^\prime}$ is defined either around the electron like or hole like
Fermi Surface;  whereas
the third term corresponds to inter-band type pairing interaction. All these terms are 
considered to arrive at the mean field Hamiltonian (\ref{ham}). As mention earlier we will solve this Hamiltonian (\ref{ham}) in three different scenarios.

\subsection{Intra-band pairing}

When only intra-band pairing terms are considered in the Hamiltonian (\ref{ham}), 
we obtained the gap equations:
\begin{eqnarray}
\label{intra-sdw}
\Delta_{SDW} = U_{SDW} \sum_{k} \left(\frac{\Delta_{SC}^{e^-}}{E_{k}^{e^{-}}}\right)
\tanh\left(\frac{\beta E_{k}^{e^{-}}}{2}\right)+\left(\frac{\Delta_{SC}^{e^+}}{E_{k}^{e^{+}}}
\right) \nonumber \\  
\times \tanh\left(\frac{\beta E_{k}^{e^{+}}}{2}\right)
+\left(\frac{\Delta_{SC}^{h^-}}{E_{k}^{h^{-}}}\right)
\tanh\left(\frac{\beta E_{k}^{h^{-}}}{2}\right) \nonumber \\ +
\left(\frac{\Delta_{SC}^{h^+}}{E_{k}^{h^{+}}}\right)\tanh\left(\frac{\beta E_{k}^{h^{+}}}
{2}\right) 
\end{eqnarray}

\begin{eqnarray}
\label{intra-odw}
\Delta_{ODW}=V_{ODW}\sum_{k} \frac{\Delta_{ODW}}{E_{k}^{e^{-}}}
\tanh\left(\frac{\beta E_{k}^{e^{-}}}{2}\right)+\frac{\Delta_{ODW}}{E_{k}^{e^{+}}} \nonumber 
\\
\times \tanh\left(\frac{\beta E_{k}^{e^{+}}}{2}\right)+ 
\frac{\Delta_{ODW}}{E_{k}^{h^{-}}}\tanh\left(\frac{\beta E_{k}^{h^{-}}}{2}\right) \nonumber \\
+\frac{\Delta_{ODW}}{E_{k}^{h^{+}}}\tanh\left(\frac{\beta E_{k}^{h^{+}}}{2}\right)
\end{eqnarray}
\begin{eqnarray}
\label{intra-sce}
\Delta_{SC}^{e}=\sum_{k^{'}}V_{kk^{'}}^e\left\{ \frac{\Delta_{SC}^{e-}}{E_{k^{'}}^{e^{-}}}\tanh\frac{\beta E_{k^{'}}^{e^{-}}}{2}-\frac{\Delta_{SC}^{e+}}{E_{k^{'}}^{e^{+}}}\tanh\frac{\beta E_{k^{'}}^{e^{+}}}{2}\right\} 
\end{eqnarray}
\begin{eqnarray}
\label{intra-sch}
\Delta_{SC}^{h}=\sum_{k^{'}}V_{kk^{'}}^h\left\{ \frac{\Delta_{SC}^{h-}}{E_{k^{'}}^{h^{-}}}\tanh\frac{\beta E_{k^{'}}^{h^{-}}}{2}-\frac{\Delta_{SC}^{h+}}{E_{k^{'}}^{h^{+}}}\tanh\frac{\beta E_{k^{'}}^{h^{+}}}{2}\right\} 
\end{eqnarray}
where the quasi particle energies in equations (\ref{intra-sdw},\ref{intra-odw},\ref{intra-sce},\ref{intra-sch})
 are obtained as,
\begin{eqnarray}
E_{k}^{e^{\pm}}&=&\pm\sqrt{\left(\varepsilon_{k}^{e}\right)^{2}+\left(\sqrt{\Delta_{SDW}^{2}
+\Delta_{ODW}^{2}}\pm\Delta_{SC}^{e}\right)^{2}} \nonumber \\
&=& \pm\sqrt{\left(\varepsilon_{k}^{e}\right)^{2}+\left(\Delta_{SC}^{e^{\pm}}\right)^{2}} 
\nonumber \\ 
E_{k}^{h^{\pm}}&=&\pm\sqrt{\left(\varepsilon_{k}^{h}\right)^{2}+\left(\sqrt{\Delta_{SDW}^{2}
+\Delta_{ODW}^{2}}\pm\Delta_{SC}^{h}\right)^{2}} \nonumber \\
&=&\pm\sqrt{\left(\varepsilon_{k}^{h}\right)^{2}+\left(\Delta_{SC}^{h^{\pm}}\right)^{2}}
\end{eqnarray}
An effective gap around the electron Fermi Surface ($\Delta_{SC}^{e^{\pm}}$) appears in the 
electronic band ($\varepsilon_{k}^{e}$) where as the same around the hole Fermi Surface ($\Delta_{SC}^{h^{\pm}}$) appears in the hole band ($\varepsilon_{k}^{h}$). In the presence of SDW and ODW orders the two SC orders $\Delta_{SC}^{e/h}(k)$ (given by equation (4,5)) are still coupled through the equations (2,3) appearing in the quasi-particle energies (6). To note that the SDW, ODW orders are inter-band in nature and thus even in intra-band pairing picture both the $\Delta_{SC}^{e/h}(k)$ orders have inter-band effect. In the intra-band picture however, the self-consistent solutions of the gap equations results in two 
SC gaps which vanish at two distinctly different T$_c$s \cite{AIP}. Such a picture would result in two 
specific heat jumps below $T_c$. These features do not support the well known ARPES data \cite{Ding}, and hence excluded from rest of our calculations. In the limiting case of vanishing $\Delta_{SDW}$, $\Delta_{ODW}$, the SC gap equations take usual BCS form,
\begin{eqnarray}
\Delta_{SC}^{e}=\sum_{k^{'}}V_{kk^{'}}^e\frac{2\Delta_{SC}^e}{E_{k^{'}}^{e}}\tanh\left(\frac{\beta E_{k^{'}}^{e}}{2}\right)
\end{eqnarray}
\begin{eqnarray}
\Delta_{SC}^{h}=\sum_{k^{'}}V_{kk^{'}}^h\frac{2\Delta_{SC}^h}{E_{k^{'}}^{h}}\tanh\left(\frac{\beta E_{k^{'}}^{h}}{2}\right)
\end{eqnarray}
Where the quasi particle energies are given as,
\begin{eqnarray}
E_{k}^{e}=\pm\sqrt{\left(\varepsilon_{k}^{e}\right)^{2}+\left(\Delta_{SC}^{e}\right)^{2}} \nonumber \\
E_{k}^{h}=\pm\sqrt{\left(\varepsilon_{k}^{h}\right)^{2}+\left(\Delta_{SC}^{h}\right)^{2}}
\end{eqnarray}

\subsection{Inter-band pairing}

We also obtain the gap equations in the inter-band pairing only, 
such gap equations take very similar form as
that of in ref \cite{two-band-prb}; temperature dependence of those results in a single T$_c$. 

\begin{eqnarray}
&&\Delta_{SDW} = U_{SDW}\sum_{k} \left(\frac{\Delta_{SC}^{e^-}}{E_{k}^{e^{-}}}\right)
\tanh\left(\frac{\beta E_{k}^{e^{-}}}{2}\right)+\left(\frac{\Delta_{SC}^{e^+}}{E_{k}^{e^{+}}}
\right) \nonumber \\&&
\times \tanh\left(\frac{\beta E_{k}^{e^{+}}}{2}\right)
+\left(\frac{\Delta_{SC}^{h^-}}{E_{k}^{h^{-}}}\right)
\tanh\left(\frac{\beta E_{k}^{h^{-}}}{2}\right)+ \nonumber \\&&
\left(\frac{\Delta_{SC}^{h^+}}{E_{k}^{h^{+}}}\right)\tanh\left(\frac{\beta E_{k}^{h^{+}}}
{2}\right)
\end{eqnarray}
\begin{eqnarray}
&&\Delta_{ODW}=V_{ODW}\sum_{k} \frac{\Delta_{ODW}}{E_{k}^{e^{-}}}
\tanh\left(\frac{\beta E_{k}^{e^{-}}}{2}\right)+\frac{\Delta_{ODW}}{E_{k}^{e^{+}}} \nonumber \\&&
\times \tanh\left(\frac{\beta E_{k}^{e^{+}}}{2}\right)+ 
\frac{\Delta_{ODW}}{E_{k}^{h^{-}}}\tanh\left(\frac{\beta E_{k}^{h^{-}}}{2}\right)+ \nonumber \\&&
~~~~~\frac{\Delta_{ODW}}{E_{k}^{h^{+}}}\tanh\left(\frac{\beta E_{k}^{h^{+}}}{2}\right)
\end{eqnarray}
\begin{eqnarray}
\label{inter-sce}
\Delta_{SC}^{e}=\sum_{k^{'}}V_{kk^{'}}^e\left\{ \frac{\Delta_{SC}^{h-}}{E_{k^{'}}^{e^{-}}}\tanh\frac{\beta E_{k^{'}}^{e^{-}}}{2}-\frac{\Delta_{SC}^{h+}}{E_{k^{'}}^{e^{+}}}\tanh\frac{\beta E_{k^{'}}^{e^{+}}}{2}\right\} 
\end{eqnarray}
\begin{eqnarray}
\label{inter-sch}
\Delta_{SC}^{h}=\sum_{k^{'}}V_{kk^{'}}^h\left\{ \frac{\Delta_{SC}^{e-}}{E_{k^{'}}^{h^{-}}}\tanh\frac{\beta E_{k^{'}}^{h^{-}}}{2}-\frac{\Delta_{SC}^{e+}}{E_{k^{'}}^{h^{+}}}\tanh\frac{\beta E_{k^{'}}^{h^{+}}}{2}\right\} 
\end{eqnarray}
where the quasi particle energies are calculated as,
\begin{eqnarray}
E_{k}^{e^{\pm}}&=&\pm\sqrt{\left(\varepsilon_{k}^{e}\right)^{2}+\left(\sqrt{\Delta_{SDW}^{2}+
\Delta_{ODW}^{2}}\pm\Delta_{SC}^{h}\right)^{2}}  \nonumber \\ 
&=& \pm\sqrt{\left(\varepsilon_{k}^{e}\right)^{2}+\left(\Delta_{SC}^{h^{\pm}}\right)^{2}} \nonumber \\
E_{k}^{h^{\pm}}&=&\pm\sqrt{\left(\varepsilon_{k}^{h}\right)^{2}+\left(\sqrt{\Delta_{SDW}^{2}+
\Delta_{ODW}^{2}}\pm\Delta_{SC}^{e}\right)^{2}} \nonumber \\ 
&=& \pm\sqrt{\left(\varepsilon_{k}^{h}\right)^{2}+\left(\Delta_{SC}^{e^{\pm}}\right)^{2}}
\end{eqnarray}

Equation(14) may be contrasted with that of the (6). Unlike the previous case of intra-band 
pairing, in the inter-band picture the effective gap ($\Delta_{SC}^{h^{\pm}}$) which involves 
SC gap around the hole FS, appears in the electronic band ($\varepsilon_{k}^{e}$). On the 
other hand, the effective gap ($\Delta_{SC}^{e^{\pm}}$) appears in the hole band 
($\varepsilon_{k}^{h}$) involves $\Delta_{SC}^{e}$, the SC gap around the electronic FS. 
Such nature of quasi-particles lead to several unusual properties like large BCS Characteristic 
ratio, identical transition temperatures to multi-gaps, their thermal behaviours and in general 
does not follow weak-coupling behaviours. In the limiting case of vanishing $\Delta_{SDW}$, 
$\Delta_{ODW}$ the SC gap equations take forms as,
\begin{eqnarray}
\label{sce-inter-sdw-odw0}
\Delta_{SC}^{e}=\sum_{k^{'}}V_{kk^{'}}^e\frac{2\Delta_{SC}^h}{E_{k^{'}}^{e}}\tanh\frac{\beta E_{k^{'}}^{e}}{2}
\end{eqnarray}
\begin{eqnarray}
\label{sch-inter-sdw-odw0}
\Delta_{SC}^{h}=\sum_{k^{'}}V_{kk^{'}}^h\frac{2\Delta_{SC}^e}{E_{k^{'}}^{h}}\tanh\frac{\beta E_{k^{'}}^{h}}{2}.
\end{eqnarray}
Where the quasi particle energies are given as,
\begin{eqnarray}
E_{k}^{e}=\pm\sqrt{\left(\varepsilon_{k}^{e}\right)^{2}+\left(\Delta_{SC}^{h}\right)^{2}} \nonumber \\
E_{k}^{h}=\pm\sqrt{\left(\varepsilon_{k}^{h}\right)^{2}+\left(\Delta_{SC}^{e}\right)^{2}}.
\end{eqnarray}

The gap equations (\ref{sce-inter-sdw-odw0}, \ref{sch-inter-sdw-odw0}) may be contrasted with that of the reference \cite{two-band-prb}.
In ref.[19] the gap equations have slightly different form than that of equations (15,16) in our work.
The difference appears in the form of quasi-particle energies, $E_{k}^{1,2} = \sqrt{{(\varepsilon_{k}^{1,2}-\mu)}^2+ \Delta_{1,2}^{2}}$ in ref [19], 
in contrast to $E_{k}^{e/h}$ [given in equations (17)]. The reason for this difference is
that no nesting between $\epsilon_k^1$ and $\epsilon_k^2$
are considered in work [19]. Also there is no considerations on influence of sign-changing superconducting order parameter (SCOP). 
Given the fact that Fe-based superconductors do show evidence of nesting, sign changing of SC-order parameter these considerations are essential. This has caused difference between our equations (15,16) and that of ref[19].
In our work, 
the results in the equations (15,16) include
consideration of inter-band nesting and sign changing effect of the SCOP
($\varepsilon_{k+Q}^{e}=-\varepsilon_{k}^{h}$, $\varepsilon_{k+Q}^{h}=-\varepsilon_{k}^{e}$ and
$\Delta_{SC}^{e}(k+Q)=-\Delta_{SC}^{h}$, $\Delta_{SC}^{h}= -\Delta_{SC}^{e}$) 
from electron like FS to the hole like FS and vice versa. 
By construction of the gap equations in this subsection, vanishing or finite magnitude of any 
of the SC-gaps $\Delta_{sc}^{e/h}$ ensures the same for the other gap $\Delta_{sc}^{h/e}$.
This is precisely the reason for a single $T_c$ in the inter-band picture and such pairing 
interaction is an essential feature in Fe-based materials. 

However, in a multi-band system like Fe-based materials intra band pairing cannot be neglected. Moreover, the thermal variation of the specific heat jump when computed based on purely inter-band pairing does not follow the $\Delta C/T_c \propto T_c^2$ form. 
In the combined intra-inter band pairing mechanism the BCS characteristic ratio, 
specific heat results resemble with experimentally observed one. 
Our findings of large values of BCS characteristic ratio is a consequence of the strong
inter band pairing. These findings not only further asserts some of the findings of the earlier 
work that the BCS theory for such superconductors is not the weak-coupling 
limit of the Eliashberg theory \cite{two-band-prb}, but also the fact that the present work is a more generalization of the 
same including magnetic, orbital orders as applicable to Fe based systems.

\subsection{Intra-Inter band pairing}

More appropriate picture that describes Fe-based superconductors, may be intra-inter band pairing. In that case all the terms of the (\ref{ham}) are to be considered. Together with the intra and inter band nature of SC-pairing interaction, the above Hamiltonian (\ref{ham}) also 
have the ability to handle sign changing as well as no-sign-changing SCOPs. In the two 
cases the Hamiltonian takes two different forms which when solved leads to two different 
set of eigenvalues namely, for sign-changing OPs,

\begin{eqnarray}
\label{quasi-sign}
E_{k}^{e^{\pm}}&=&\pm\sqrt{\left(\varepsilon_{k}^{e}\right)^{2}+\left(\sqrt{\Delta_{SDW}^{2}
+\Delta_{ODW}^{2}}\pm\Delta_{SC}(k)\right)^{2}}  \nonumber \\ &=&
\pm\sqrt{\left(\varepsilon_{k}^{e}\right)^{2}+\left(\Delta_{SC}^{\pm}(k)\right)^{2}} \nonumber\\
E_{k}^{h^{\pm}}&=&\pm\sqrt{\left(\varepsilon_{k}^{h}\right)^{2}+\left(\sqrt{\Delta_{SDW}^{2}
+\Delta_{ODW}^{2}}\pm\Delta_{SC}(k)\right)^{2}} \nonumber \\ &=& 
\pm\sqrt{\left(\varepsilon_{k}^{h}\right)^{2}+\left(\Delta_{SC}^{\pm}(k)\right)^{2}},
\end{eqnarray}
and for no-sign-changing OPs,
\begin{eqnarray}
\label{quasi-no-sign}
E_{k}^{e}=\pm\sqrt{\left(\varepsilon_{k}^{e}\right)^{2}+\Delta_{SDW}^{2}+\Delta_{ODW}^{2}+\left(\Delta_{SC}(k)\right)^{2}} \nonumber\\
E_{k}^{h}=\pm\sqrt{\left(\varepsilon_{k}^{h}\right)^{2}+\Delta_{SDW}^{2}+\Delta_{ODW}^{2}+\left(\Delta_{SC}(k)\right)^{2}}
\end{eqnarray}
when intra and inter-band pairing are treated on an equal footing.

We also obtain and solve the gap equations involving various orders to calculate specific heat. The gap equations in the sign-changing OP scenario are given as below.

\begin{eqnarray}
\label{SDW-sign}
&&\Delta_{SDW}=U_{SDW}\sum_{k}\frac{\Delta_{SC}^-}{E_{k}^{e^{-}}}
\tanh\left(\frac{\beta E_{k}^{e^{-}}}{2}\right)+ \frac{\Delta_{SC}^+}{E_{k}^{e^{+}}} \nonumber \\
&& \times \tanh\left(\frac{\beta E_{k}^{e^{+}}}{2}\right) 
+\left(\frac{\Delta_{SC}^-}{E_{k}^{h^{-}}}\right)
\tanh\left(\frac{\beta E_{k}^{h^{-}}}{2}\right)+ \nonumber \\ &&
\left(\frac{\Delta_{SC}^+}{E_{k}^{h^{+}}}\right)\tanh\left(\frac{\beta E_{k}^{h^{+}}}{2}\right) 
\end{eqnarray}
\begin{eqnarray}
\label{ODW-sign}
&& \Delta_{ODW}=U_{ODW}\sum_{k} \frac{\Delta_{ODW}}{E_{k}^{e^{-}}}
\tanh\left(\frac{\beta E_{k}^{e^{-}}}{2}\right)+\frac{\Delta_{ODW}}{E_{k}^{e^{+}}}
\nonumber \\ &&
\times \tanh\left(\frac{\beta E_{k}^{e^{+}}}{2}\right) 
+\frac{\Delta_{ODW}}{E_{k}^{h^{-}}}\tanh\left(\frac{\beta E_{k}^{h^{-}}}{2}\right)+
\nonumber \\ && 
\frac{\Delta_{ODW}}{E_{k}^{h^{+}}}\tanh\left(\frac{\beta E_{k}^{h^{+}}}{2}\right) 
\end{eqnarray}
\begin{eqnarray}
\label{SCe-sign}
\Delta_{SC}^{e}(k)=\sum_{k^\prime} V_{k k^\prime}^{e}\left\{ 
\frac{\Delta_{SC}^{^{-}}}
{E_{k^\prime}^{e^{-}}}\tanh\frac{\beta E_{k^\prime}^{e^{-}}}{2}-
 \frac{\Delta_{SC}^{^{+}}}{E_{k^\prime}^{e^{+}}}\tanh
\frac{\beta E_{k^\prime}^{e^{+}}}{2}\right\} 
\end{eqnarray}
\begin{eqnarray}
\label{SCh-sign}
\Delta_{SC}^{h}(k)=\sum_{k^\prime}V_{k k^\prime}^{h}\left\{ \frac{\Delta_{SC}^{^{-}}
}{E_{k^\prime}^{h^{-}}}\tanh\frac{\beta E_{k^\prime}^{h^{-}}}{2}-
\frac{\Delta_{SC}^{^{+}}}{E_{k^\prime}^{h^{+}}}\tanh
\frac{\beta E_{k^\prime}^{h^{+}}}{2}\right\} 
\end{eqnarray}
For no-sign-changing OP symmetries we have obtained,
\begin{eqnarray}
\label{SDW-no-sign}
\frac{\Delta_{SDW}}{2 U_{SDW}}=\sum_{k}\left\{ \frac{\tilde{\Delta}}{E_{k}^{e}}\tanh\frac{\beta E_{k}^{e}}{2}+\frac{\tilde{\Delta}}{E_{e}^{h}}\tanh\frac{\beta E_{k}^{h}}{2}\right\} 
\end{eqnarray}
\begin{eqnarray}
\label{ODW-no-sign}
\frac{\Delta_{ODW}}{2 U_{ODW}}=\sum_{k}\left\{ \frac{\Delta_{ODW}}{E_{k}^{e}}\tanh\frac{\beta E_{k}^{e}}{2}+\frac{\Delta_{ODW}}{E_{k}^{h}}\tanh\frac{\beta E_{k}^{h}}{2}\right\} 
\end{eqnarray}
\begin{eqnarray}
\label{SCe-no-sign}
\Delta_{SC}^{e}=\sum_{k^{'}}V_{kk^{'}}^e\frac{2\Delta_{SC}}{E_{k^{'}}^{e}}\tanh\left(\frac{\beta E_{k^{'}}^{e}}{2}\right)
\end{eqnarray}
\begin{eqnarray}
\label{SCh-no-sign}
\Delta_{SC}^{h}=\sum_{k^{'}}V_{kk^{'}}^h\frac{2\Delta_{SC}}{E_{k^{'}}^{h}}\tanh\left(\frac{\beta E_{k^{'}}^{h}}{2}\right)
\end{eqnarray}
For the set of gap equations (\ref{SDW-sign}, \ref{ODW-sign}, \ref{SCe-sign}, 
\ref{SCh-sign}) the quasi-particle energies involved are given by (\ref{quasi-sign}) whereas
for the set of gap equations (\ref{SDW-no-sign}, \ref{ODW-no-sign}, \ref{SCe-no-sign}, 
\ref{SCh-no-sign}) the quasi-particle energies involved are given by (\ref{quasi-no-sign}).

Therefore, we solve these four gap equations numerically following the procedure as
in \cite{Ghosh} for different allowed pairing symmetries like 
d$_{x^2-y^2}$+s$_{x^2+y^2}$ / s$^{\pm}$, s$_{xy}$ 
that changes sign between the electron and hole like Fermi Surface and isotropic s-wave for
no sign changing OP, in the combined intra-inter band pairing mechanism. 
Variation of SCOPs (energy gap) with temperature, as obtained 
from the four coupled equations, can be used to calculate 2$\Delta_{SC}/k_BT_c$ as well as 
specific heat as a function of temperature. Specific heat can be obtained from the 
electronic entropy which is defined as:
\begin{figure}[ht]
%\epsfxsize=4.0truein
%\epsfysize=3.5truein
%\epsffile{inter.eps}
%\centering
\includegraphics[width=8.5cm]{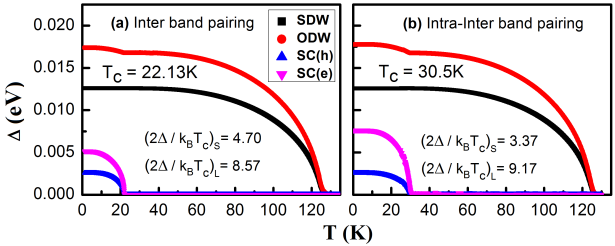}
\caption{Temperature variation of various OPs like SDW (in black), ODW (in red)
 and superconductivity (electron FS violet, Hole FS blue) 
in only inter-band {\bf (left)} and intra-inter {\bf (right)} band pairing for hole doped ($\mu=1.42$) system having d+s pairing 
symmetry. Characteristics ratios and SC $T_c$s are indicted in the figure.}
\label{inter}
\end{figure}

\begin{eqnarray}
S_{es}=-2k_B\sum_k[(1-f_k)\ln (1 - f_k) + f_k ln f_k]
\end{eqnarray}
where $f_k = (1 + e^{\beta E_k})^{-1}$ is the Fermi function and $\beta = 1/k_BT$. Electronic specific heat can be found using the relation C= -$\beta \frac{dS_{es}}{d\beta}$
\begin{figure}[ht]
%\epsfxsize=4.0truein
%\epsfysize=3.5truein
%\epsffile{d+s_e.eps}
\centering
\includegraphics[width=8.5cm]{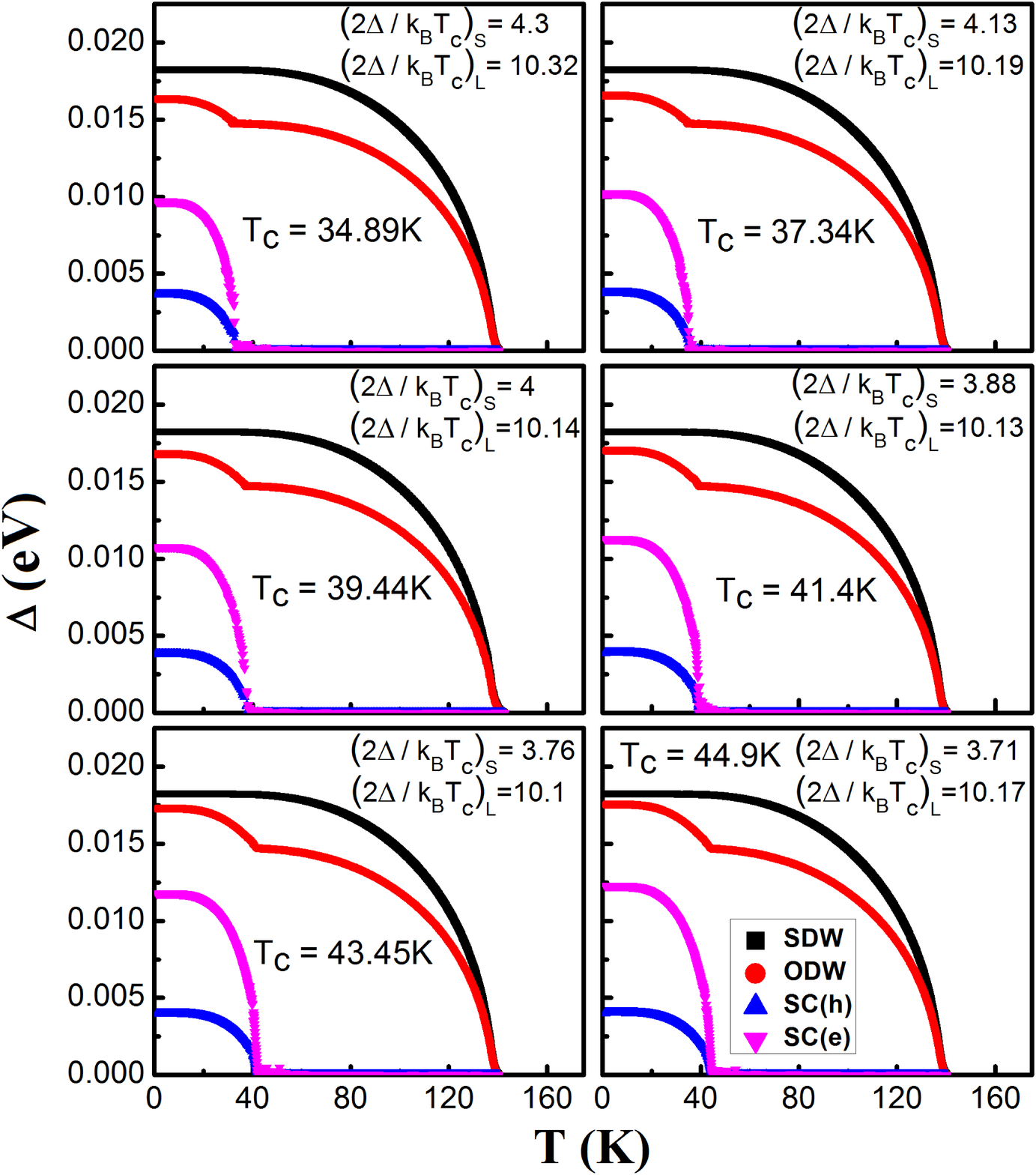}
\caption{Temperature variation of various OPs like SDW (in black), ODW (in red) and superconductivity (electron FS
 violet, Hole FS blue) 
for different 122 FePn having different T$_c$s for electron doped ($\mu=1.53$) system having d+s pairing 
symmetry. Variation of BCS characteristic ratio with different T$_c$s are indicated.}
\label{d+s-e-gap}
\end{figure}

Different pairing symmetries are imposed in SCOPs which 
significantly modifies the temperature variation of all the OPs as they 
are coupled with each other (see for details in the next section). 
Doping (electron or hole) is controlled by chemical potential $\mu$.
Behaviour of the electronic specific heat particularly the jump in specific heat are 
also modified depending on the pairing symmetry.

\section{Results and Discussions}

\subsection{Calculation of BCS Characteristic ratio :}
\begin{figure}[ht]
%\epsfxsize=4.0truein
%\epsfysize=3.5truein
%\epsffile{d+s_h.eps}
\centering
\includegraphics[width=8.5cm]{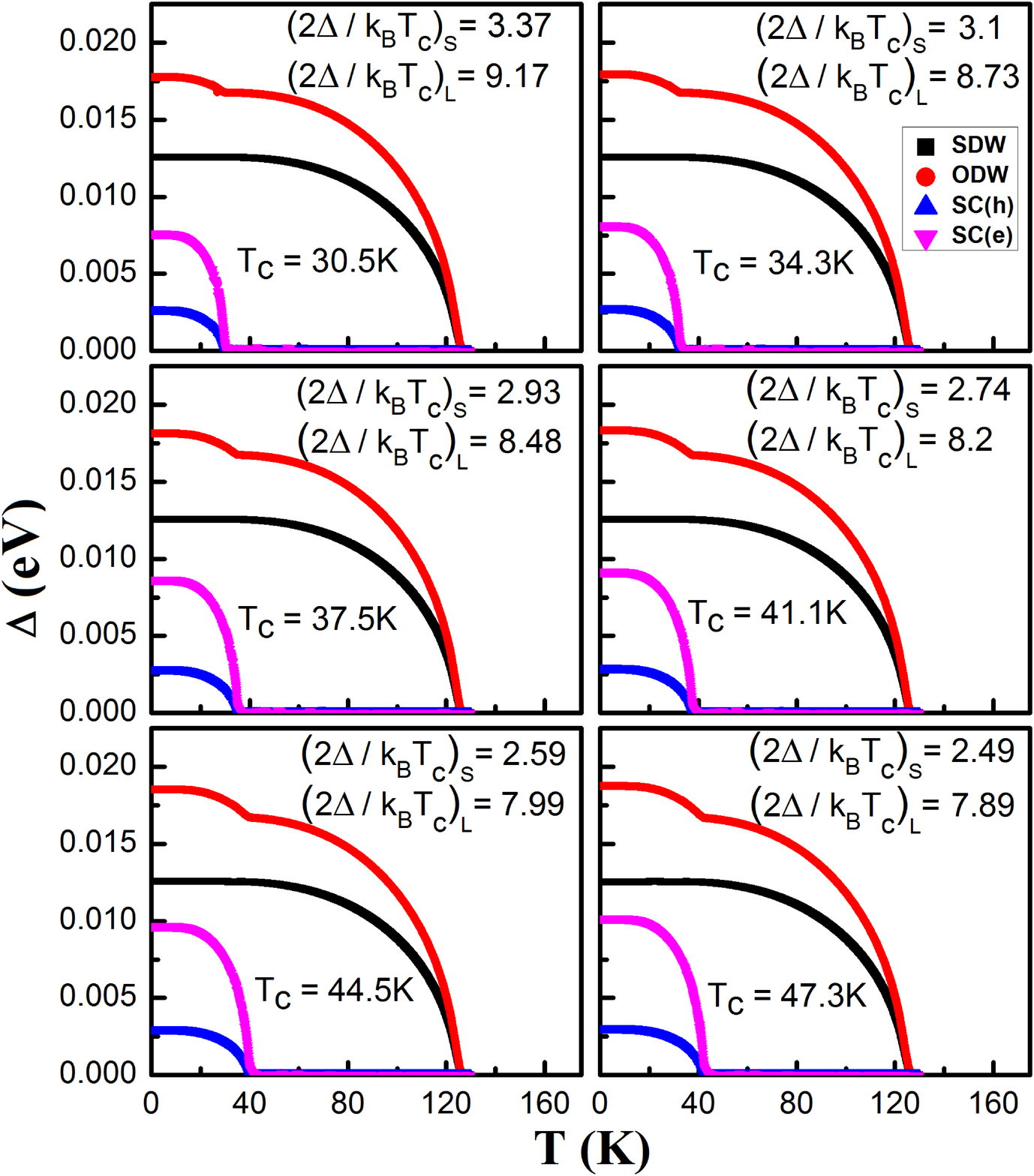}
\caption{Temperature variation of various OPs like SDW (in black), ODW (in red)
 and superconductivity (electron FS violet, Hole FS blue) 
for different 122 FePn having different T$_c$s for hole doped ($\mu=1.42$) system having d+s pairing 
symmetry. Variation of BCS characteristic ratio with different T$_c$s are indicated}
\label{d+s-h-gap}
\end{figure}
BCS characteristics ratio, is defined as $2\Delta_{SC}/k_BT_c$ , where $\Delta_{SC}$ is the 
SC gap at T=0K. Weak coupling BCS theory predicts characteristics ratio of 
conventional superconductors as 3.5. We have solved all the four coupled gap equations 
numerically for three cases (i) intra-band (ii) inter-band and (iii) intra-inter band pairing 
on an equal footing to get different OPs (SDW, ODW, SC around electron and hole FS)
 as a function of temperature. In case of intra-band pairing we got two different $T_c$s for 
two SCOPs (electron and hole band) as reported earlier \cite{AIP}. As this 
behaviour is not consistent with the experiments, other two possibilities (inter-band and 
intra-inter band pairing) are examined thoroughly. Temperature dependence of various order 
parameters (SDW, ODW, SC around electron and hole FS) for inter-band and intra-inter band 
pairing are shown in FIG.\ref{inter}(a) and FIG.\ref{inter}(b) respectively for d+s pairing symmetry (all other conditions remain identical for both the cases).\begin{figure}[ht]
%\epsfxsize=4.0truein
%\epsfysize=3.5truein
%\epsffile{sxy_h.eps}
\centering
\includegraphics[width=8.5cm]{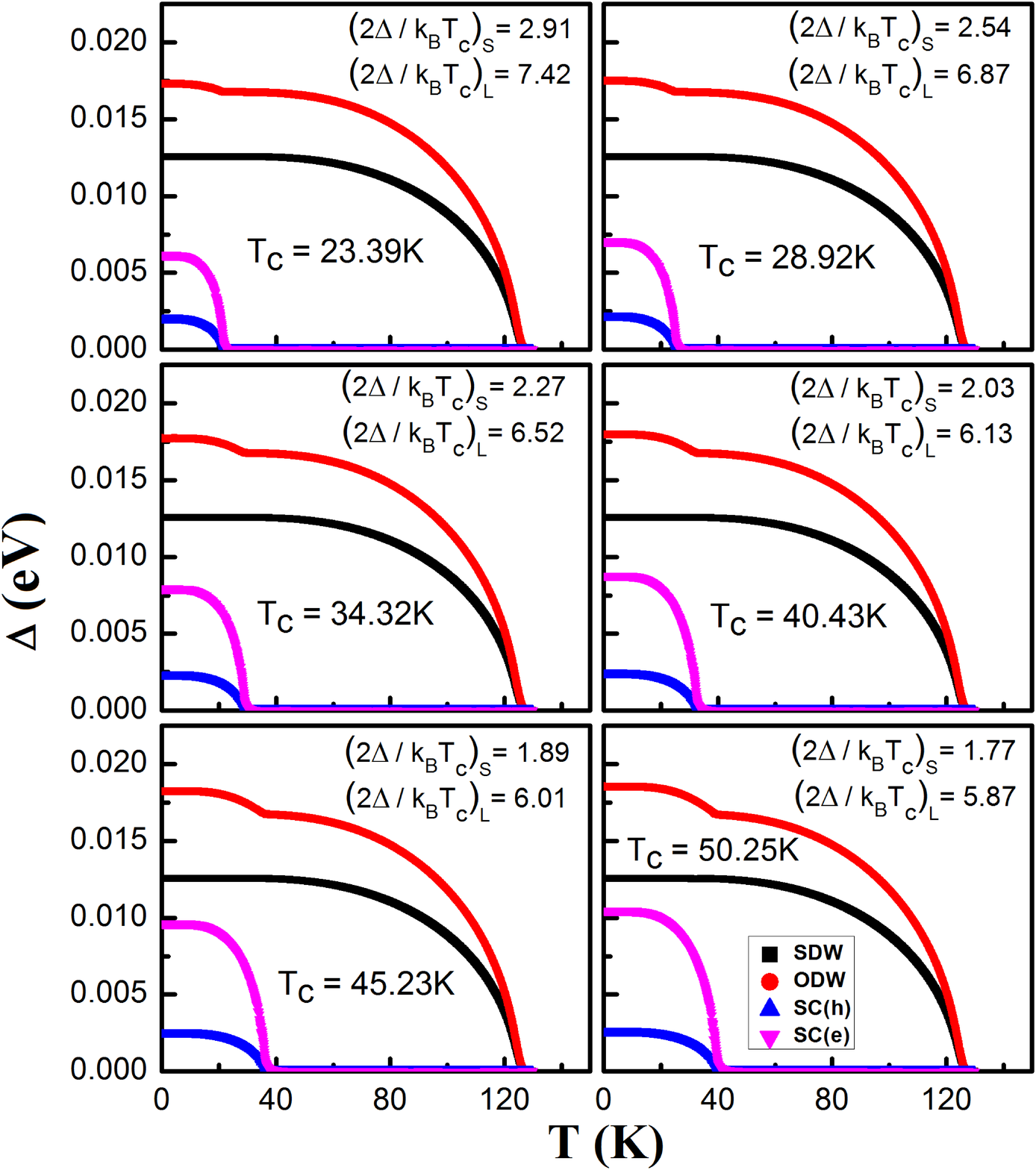}
\caption{Temperature variation of various OPs like SDW (in black), ODW (in red)
 and superconductivity (electron FS violet, Hole FS blue) 
for different 122 FePn having different T$_c$s for hole doped ($\mu=1.42$) system having s$_{xy}$ pairing 
symmetry. Variation of BCS characteristic ratio with different T$_c$s are indicated}
\label{sxy-h-gap}
\end{figure} A closer look to the SC gap equations [sec IIB, equations(\ref{inter-sce},\ref{inter-sch})] in the purely inter-band picture indicates the following. If at a given temperature and doping $\Delta_{sc}^e$ becomes zero (or finite) then it simultaneously make $\Delta_{sc}^h$ also zero (or finite). That is both the SCOPs either exist or does not exist, 
ensuring simultaneous opening up of both the gaps.
 Since respective gaps are opened to their partner's band density of states, there is a competition between it. So growth of both of the gaps are competitive leading to large $2\Delta_{SC}/k_BT_c$ ratio. In the combined intra-inter band picture however; the pairing strength contribution from intra-band one leads to opening up of any of the gaps slightly higher in temperature leading to higher $T_c$. This also leads to larger growth of the gaps at the lower temperatures leading $\Delta^{large}(0)/\Delta^{small}(0)$ towards 3. This is also the reason for moderated $2\Delta_{SC}/k_BT_c$ ratio in this picture. Temperature dependencies are very similar in both the cases, but superconductivity is more favoured in the combined intra-inter band case and $T_c$ is smaller in inter-band only pairing compared to that of intra-inter band picture. The zero temperature gap ratio (large to small) in the inter-band only pairing is slightly less than 2 whereas that in the intra-inter band picture is greater than 2 ($\sim 2.5-3$) \cite{PRB2014}. The later scenario matches with the experimental scenario much better. FIG.\ref{d+s-e-gap}, FIG.\ref{d+s-h-gap}, FIG.\ref{sxy-h-gap}, FIG.\ref{iso} shows the temperature variation of SDW, ODW and SCOPs for various OP symmetries of the SC state like d+s, $s_{xy}$ and isotropic s-wave considering combined intra and inter band pairing on an equal footing. In all those figures (Fig 2-6) SDW OP, ODW OP, and SCOP for electron and hole like Fermi surfaces are represented through black, red, violet and blue respectively. These thermal variations of various OPs are used to establish the influence of OP symmetries in specific heat calculations. At T=0K the SC gap (both around electron and hole FS) is maximum, we take it as $\Delta_{sc}$ (T=0). \begin{figure}[ht]
%\epsfxsize=4.0truein
%\epsfysize=3.5truein
%\epsffile{iso.eps}
\centering
\includegraphics[width=8.5cm]{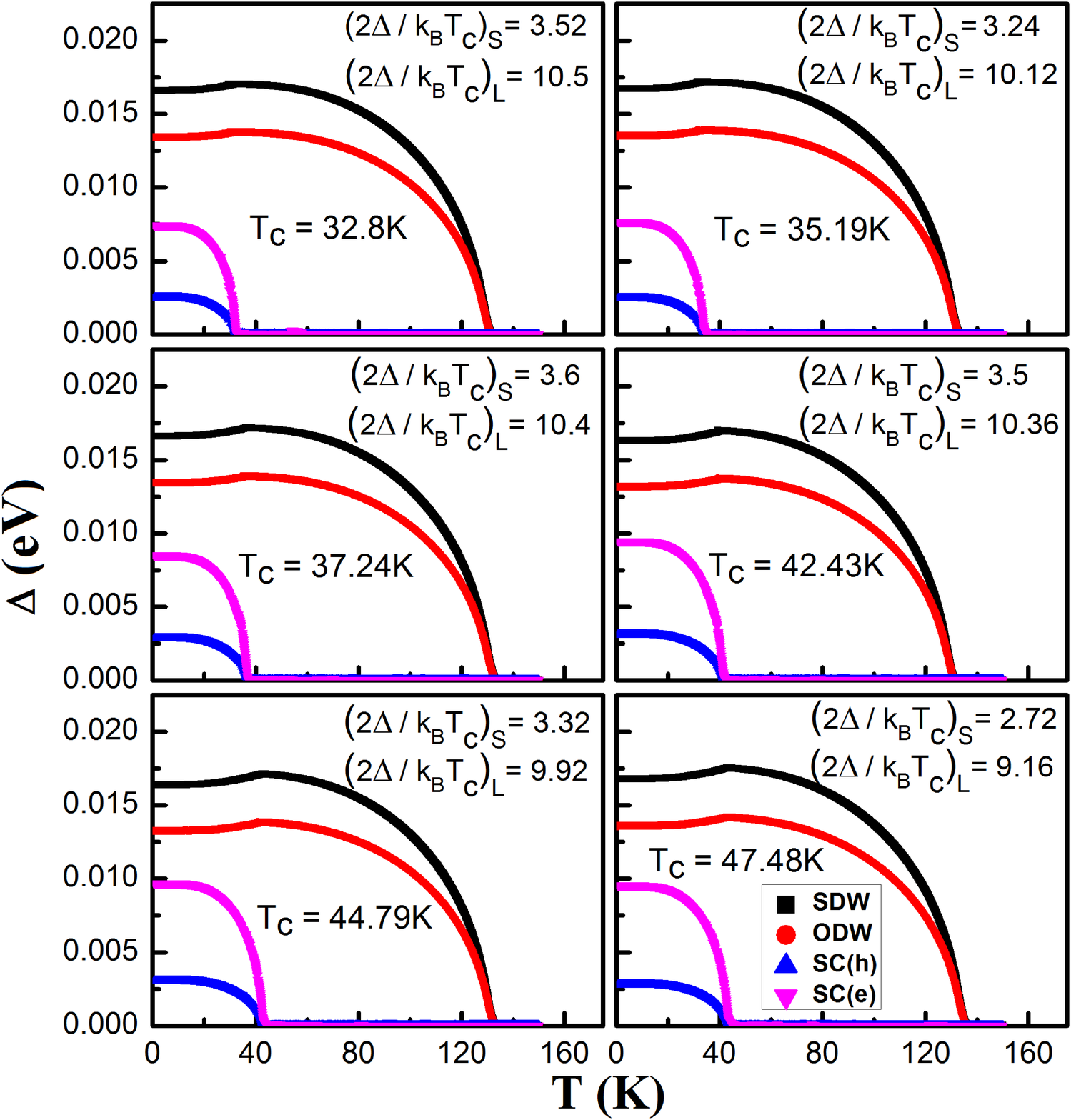}
\caption{Temperature variation of various OPs like SDW (in black), ODW (in red)
 and superconductivity (electron FS violet, Hole FS blue) 
for different 122 FePn having different T$_c$s for electron doped system having isotropic s-wave pairing 
symmetry. Variation of BCS characteristic ratio with different T$_c$s are indicated}
\label{iso}
\end{figure}As there are two energy gaps (around electron and hole like FS) we got two 
characteristics ratios one is large for $\Delta_{sc}^e$ and other one is small for 
$\Delta_{sc}^h$. The momentum averaged $\Delta_{sc}^{small}(k)$ is obtained by taking average 
of $\Delta_{sc}^h$ over the outer $\Gamma$ Fermi line of FIG.\ref{FS} whereas the momentum 
averaged $\Delta_{sc}^{large}(k)$ is obtained by taking average of $\Delta_{sc}^e$ over the 
inner $\Gamma$ Fermi line. In doing so, momentum dependence of the SCOPs for 
various pairing symmetries are considered. In each of the four cases (electron and hole doped 
d+s wave, electron doped isotropic s-wave and hole doped $s_{xy}$ pairing symmetry) we have 
found the value of large and small $2\Delta_{SC}/k_BT_c$ for different transition temperatures 
and their values are presented inside the figures for each set. Since in this work
 we are predicting properties like $2\Delta_{SC}/k_BT_c$, $\Delta C/T_c$ etc. as a function of 
  $T_c$, we need to vary $T_c$ and calculate these properties. The SC transition 
temperatures can be varied either by changing chemical potential $\mu$ (for hole and electron 
doped cases) or by modifying the effective attractive electron-electron interaction strength 
$V_0^{e/h}$ (where $V_{kk^\prime}^{e/h}$ are factorized as $V_0^{e/h}\eta_k\eta_{k^\prime}$, 
the momentum dependencies of $\eta_k$ determines the symmetry of the SCOP).
While the values of chemical potentials
  are presented in each figures 3--5, variations in $T_c$s are obtained as explained above and 
  its values are presented in each figure. Specific heats for a particular pairing symmetry are 
  calculated using the temperature dependencies of various order parameters in corresponding 
  pairing symmetries.
 In 
only inter-band scenario the value of $2\Delta_{SC}^{small}/k_BT_c$ is larger compared to that 
from the experimental observation of $\sim 2.5\pm1.5$. From our calculation we have got (in d+s 
pairing symmetry) small and large $2\Delta_{SC}/k_BT_c$ values around 4 and 10 for electron 
doped system and 3 and 9 for hole doped system respectively. Both small and large 
$2\Delta_{SC}/k_BT_c$ values are smaller in the hole doped case (around 3 and 9) which is 
also consistent with experimental results \cite{Evtushinsky}.
\begin{figure}[ht]
%\epsfxsize=4.0truein
%\epsfysize=3.5truein
%\epsffile{d+s_sph_e.eps}
\centering
\includegraphics[width=8.5cm]{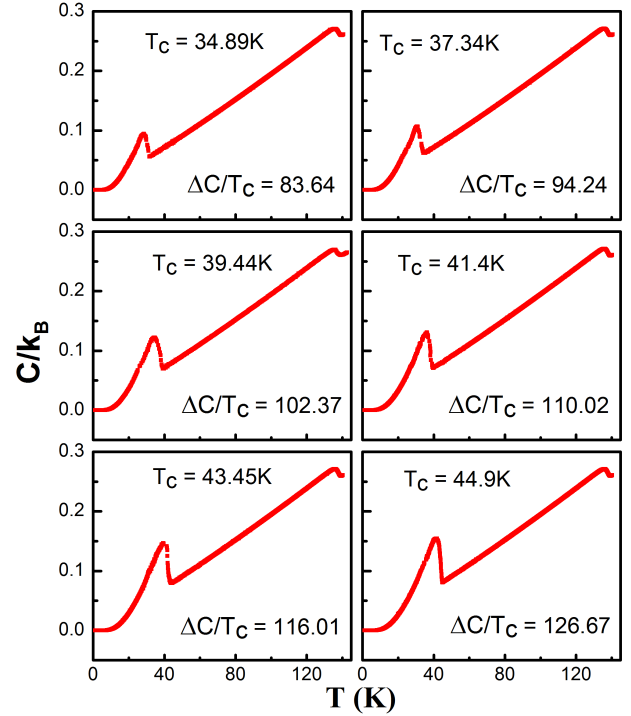}
\caption{Variation of specific heat as a function of temperature at different T$_c$ for electron doped 122 FePn systems having d+s pairing symmetry and corresponding $\Delta$C/T$_c$
 values indicated in the graph in mJ/moleK$^2$.}
\label{d+s-e-sph}
\end{figure}
\begin{figure}[ht]
%\epsfxsize=4.0truein
%\epsfysize=3.5truein
%\epsffile{d+s_sph_h.eps}
\centering
\includegraphics[width=8.5cm]{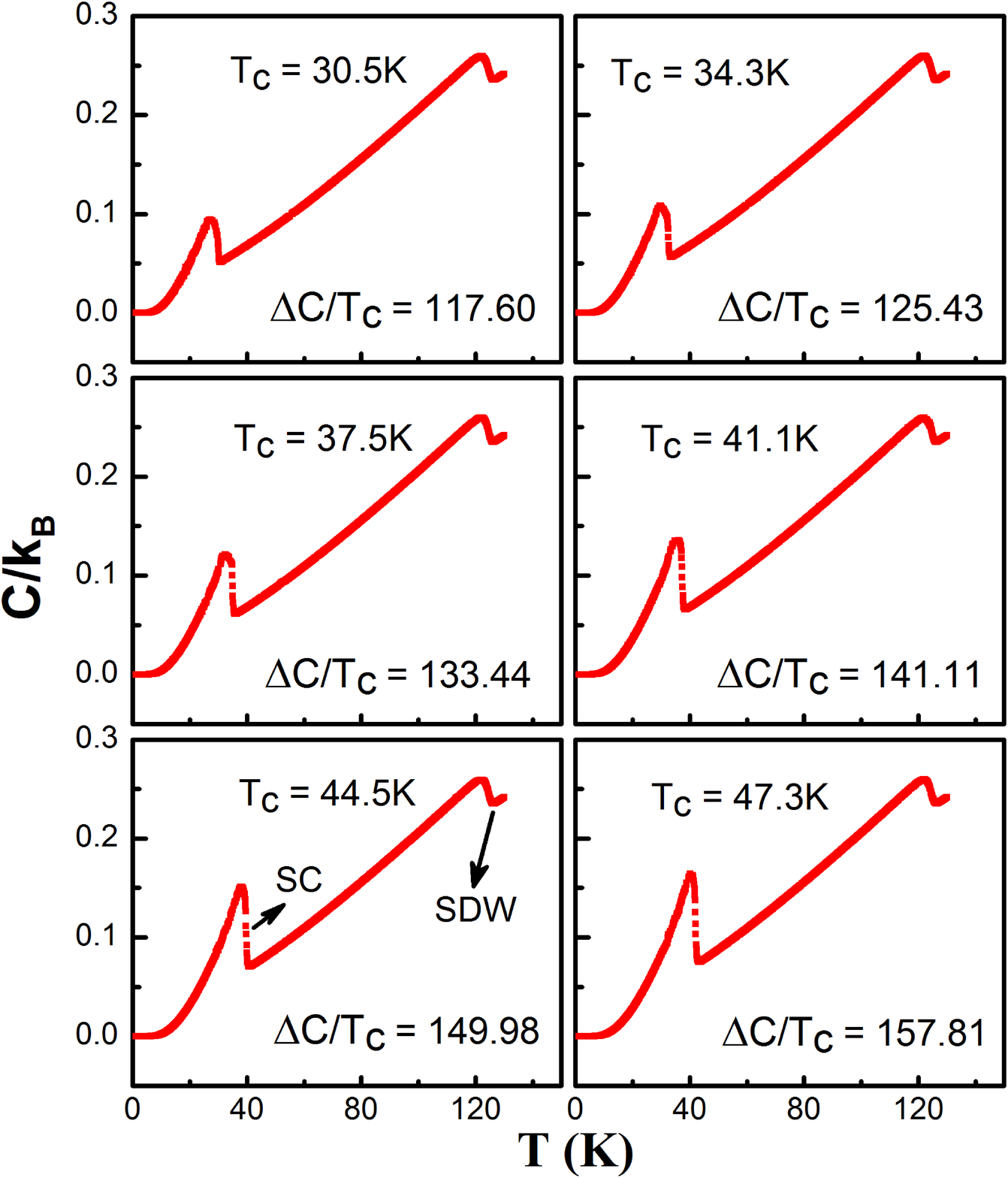}
\caption{Variation of specific heat as a function of temperature at different T$_c$ for hole
 doped 122 FePn systems having d+s pairing symmetry and corresponding $\Delta$C/T$_c$ 
values indicated in the graph in mJ/moleK$^2$. SC and SDW/ODW jumps are indicated in one of the figure.}
\label{d+s-h-sph}
\end{figure}

\subsection{Thermal variation of specific heat}

Temperature dependent specific heat is calculated using the relation mentioned above in the 
theoretical model section. In this subsection we present the behaviour of specific heat as a 
function of temperature for inter-band only and combined intra-inter band scenarios. The 
SC gap (in the inter-band picture only) around the hole FS uses the density of 
states near the electron FS and vice versa. As a result, for example,
 when $V_h^0$ is raised (which increases the Cooper pair binding around the hole FS but uses the
states around electron FS for pairing)
 to increase the 
SC $T_c$, the increment is only nominal compared to that in the intra-inter band 
picture. 
\begin{figure}[ht]
%\epsfxsize=4.0truein
%\epsfysize=3.5truein
%\epsffile{sxy_sph.eps}
\centering
\includegraphics[width=8.5cm]{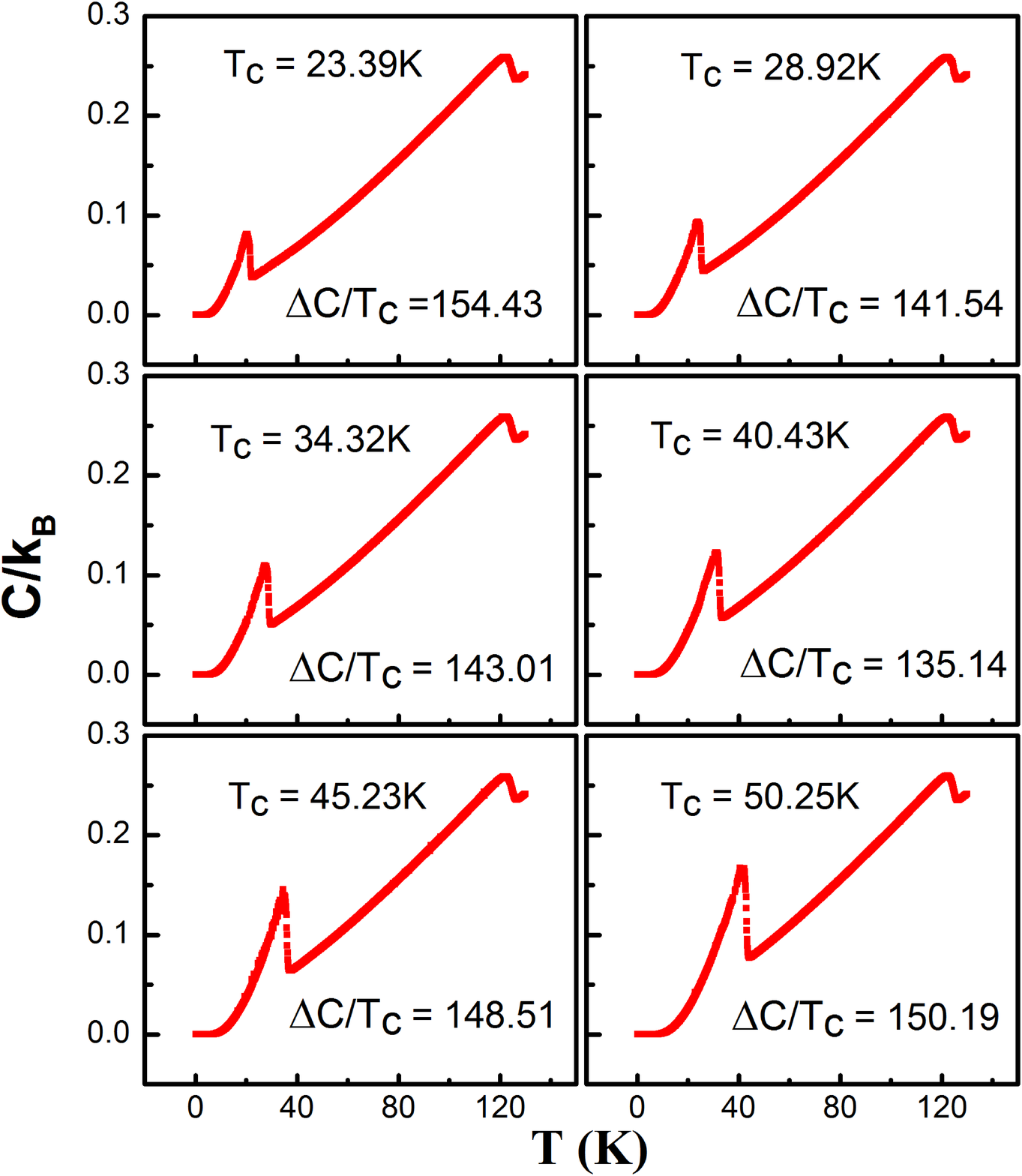}
\caption{Variation of specific heat as a function of temperature at different T$_c$ for hole
 doped systems having s$_{xy}$ pairing symmetry and corresponding $\Delta$C/T$_c$
 values indicated in the graph in mJ/moleK$^2$.}
\label{sxy-sph}
\end{figure}
This also causes a distinct difference in the temperature dependencies of specific heat. 
This in turn causes difference in $\Delta C/T_c$ vs $T_c^2$ dependence. 
FIG.\ref{all}c shows the variation 
of specific heat as a function of temperature for both only inter-band and intra-inter band 
cases. From FIG.\ref{all}c it is very clear that specific heat jump is smaller in only inter 
band picture. In intra-inter band case all the four allowed pairing symmetries are considered 
as indicated earlier. FIG.\ref{d+s-e-sph}, FIG.\ref{d+s-h-sph}, FIG.\ref{sxy-sph} and 
FIG.\ref{isos-sph} shows the variation of specific heat with temperature for d+s (electron and 
hole doped), $s_{xy}$(hole doped) and isotropic s-wave pairing (electron doped) symmetry 
respectively. In each case, we have calculated the specific heat jump $\Delta C$ at different 
$T_c$s and plotted $\Delta C/T_c$  as a function of $T_c^2$. Our calculated value of specific 
heat is in the unit of eV per 2 atoms. Most of the experimental results {\it i.e}., the value 
of specific heat are in the unit of mJ/moleK. Scaling between 
mole and atom needs to be considered in order to compare theoretical results with that of 
the experiment. For example, in 122 system that contains 5 atoms, then without concern to 
whether all the atoms has greater or lesser contribution to the Fermi level (in case of 122 
system, a mole of 122 is not considered to be consists of only two Fe atoms even though the 
contribution of density of states at Fermi level mostly comes from the Fe orbitals for these 
material) one has to multiply the value of specific heat by a factor n (n = 5 for 122 case) 
\cite{Stewart}. From these figures we clearly see that there are two jumps in the specific heat 
value, one at low temperature for SC transition ($T_c$) and other one at higher 
transition temperature for SDW and ODW.
\begin{figure}[ht]
%\epsfxsize=4.0truein
%\epsfysize=3.5truein
%\epsffile{iso_sp.eps}
\centering
\includegraphics[width=7.5cm]{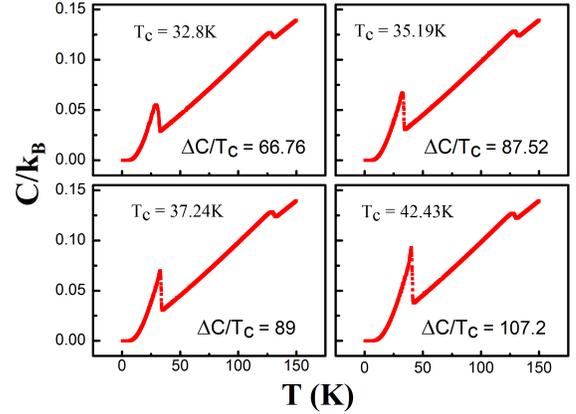}
\caption{Variation of specific heat as a function of temperature at different T$_c$ for electron
 doped systems having isotropic s-wave pairing symmetry and corresponding $\Delta$C/T$_c$
 values indicated in the graph in mJ/moleK$^2$.}
\label{isos-sph}
\end{figure}

Calculated values of $\Delta C/T_c$  with a fixed $T_c$, for electron and hole doped systems  with d+s pairing symmetry matches well with the experimental results \cite{Paglione,Gofryk}. The estimated value of $\Delta C/T_c$  for other pairing symmetries like $s_{xy}$, isotropic s-wave are not very consistent with the experimental observation. $\Delta C/T_c$ is nearly constant with $T_c$ for $s_{xy}$ pairing symmetry (see FIG.\ref{sxy-sph}). FIG.\ref{all}a and FIG.\ref{all}b shows that $\Delta C/T_c$ is proportional to $T_c^2$ [in those figure of $\Delta C/T_c$  vs $T_c^2$, theoretical data points are compared with linear curve (solid red line)] for both electron and hole doped system with d+s pairing symmetry which is consistent with the experimental findings \cite{Paglione,Kim}. For other paring symmetry the behaviour of $\Delta C/T_c$ vs $T_c^2$ is not very clear as far as our calculation is concerned but certainly it is not proportional to $T_c^2$.

\section{Summary and Conclusion}

We present within two-band model of superconductivity a detailed study of BCS characteristic ratio and electronic specific heat. To calculate the above properties we present detailed study on the temperature dependencies of various OPs, like SDW, ODW and superconductivity in the electron and hole bands. Our entire work in the present paper may be summarized as follows. Three scenarios of SC pairings are considered. (i) intra-band pairing (ii) inter-band pairing and (iii) intra-inter band pairing.
\begin{figure}[ht]
%\epsfxsize=4.0truein
%\epsfysize=3.5truein
%\epsffile{Graph.eps}
%\centering
\flushleft\includegraphics[width=9cm]{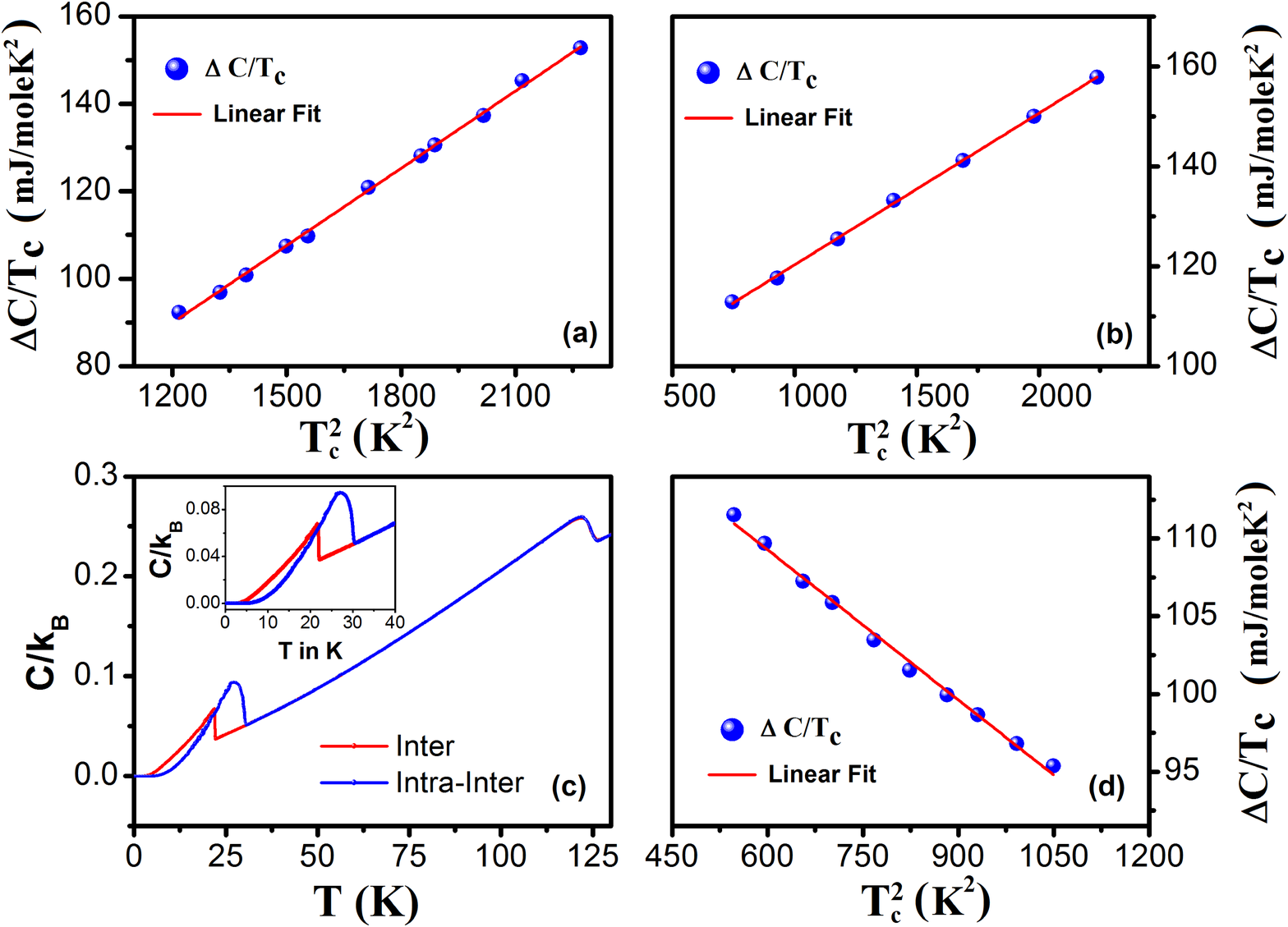}
\caption{ Variation of $\Delta$C/T$_c$  as a function of T$_{c}^2$ for (a) electron and (b) hole doped 
122 FePn systems having d+s pairing symmetry. (c) Specific heat as a function of temperature in only inter-band vs combined intra-inter band pairing. (d) Variation of $\Delta$C/T$_c$  as a function of T$_{c}^2$ for only inter-band pairing.}
\label{all}
\end{figure}
 Superconductivity within all the above scenarios are studied in presence of inter-orbital SDW and ODW order together with different allowed pairing symmetries. The intra-band pairing leads to two distinct $T_c$s and characteristic ratios similar to the weak coupling BCS theory and hence found not suitable for Fe-based superconductors , as it does not have much experimental evidence. In the solely inter-band pairing picture, single global $T_c$ is achieved and larger $2\Delta_{SC}/k_BT_c$ consistent with experimental findings are seen. This picture still suffers from drawbacks in the following (a) $\Delta^{large}/\Delta^{small}\leqslant 2$, (b) $2\Delta_{SC}^{small}/k_BT_c$ exceeds experimental findings, (c) $\Delta C/T_c\propto T_c^2$ with proportionality constant which is negative. In the third scenario with combined intra-inter band pairing all the above mentioned shortcomings are overcome. In all the above pictures coupled gap equations involving SDW, ODW and SC-gaps are presented. Nature of quasi-particle in the above three pictures are also pointed out. Specially, in the inter-band only picture nature of gap equations (in absence of magnetic and orbital orders) reproduces that of the ref.\cite{two-band-prb}. The larger value of $2\Delta_{SC}/k_BT_c$ is found to be primarily due to the presence of inter-band pairing (this includes also the conclusion of ref.\cite{two-band-prb}). Within combined intra-inter band pairing for sign changing OPs we find that the temperature dependence of specific heat jump is very different from other classes of superconductors like conventional el-ph mediated BCS superconductors, A15 compounds, high $T_c$ cuprates. We have shown that the Characteristics ratios and $\Delta C/T_c$ variation with $T_c$ matches very well with experimental findings \cite{Evtushinsky} in case of d+s pairing symmetry (for both electron and hole doped). Therefore, combined intra and inter band pairing reproduces important features from experiment. 

Finally, sign-changing $d_{x^2-y^2}+s_{x^2+y^2}$ pairing symmetry reproduces the desired 
$\Delta C/T_c$ as  function of $T_c^2$ behaviour than other pairing symmetries in the combined 
intra-inter band pairing. 
Such paring symmetry is very much consistent with the recent trends of experimental and 
theoretical research in the field \cite{Fang,Davis,Fernandes,Nakajima,Jiang,Chu}. 
The d+s pairing symmetry are consistent with the nematic phase observed in the phase diagram 
of Fe-based systems; according to this scenario the electronic ground state preserves 
the translational symmetry of the crystal but not the rotational symmetry \cite{Ming}. 
Furthermore, we have argued elsewhere \cite{Ghosh} that the d+s pairing symmetry is equivalent 
to $s^{\pm}$ symmetry in other models. We demonstrate that independent of pairing mechanism any 
theoretical model for Fe-based superconductors should contain contribution from both the 
intra and inter band pairing channels.

\section{Acknowledgements} One of us (SS) acknowledges the HBNI, RRCAT for financial support and encouragements. We thank Dr. G. S. Lodha and Dr. P.D. Gupta for their 
encouragement in this work.
%% \bibitem[ (1)]{}


\begin{thebibliography}{9}
\bibitem{Kamihara}Y. Kamihara, T. Watanabe, M. Hirano and H. Hosono,
 J. Am. Chem. Soc. {\bf 130}(11), 3296 (2008).
\bibitem{Stewart}G. R. Stewart, Rev. Mod. Phys. {\bf 83}, 1589 (2011).
\bibitem{Chen}H. Chen, Y. Ren, Y. Qiu, W. Bao, R. H. Liu, G. Wu, T. Wu, Y. L. Xie, 
X. F. Wang, Q. Huang and X. H. Chen, Europhys. Lett. {\bf 85}, 17006 (2009).
\bibitem{Nandi S}S. Nandi, M. G. Kim, A. Kreyssig, R. M. Fernandes, D. K. Pratt, 
A. Thaler, N. Ni, S. L. Bud’ko, P. C. Canfield, J. Schmalian, R. J. McQueeney and A. I. Goldman, 
Phys. Rev. Lett. {\bf 104}, 057006 (2010).
\bibitem{Tsuei}C. C. Tsuei and J. R. Kirtley, Rev. Mod. Phys. {\bf 72}(4), 969 (2000).
\bibitem{Van Harlingen}D. J. Van Harlingen, Rev. Mod. Phys. {\bf 67}(2), 515 (1995).
\bibitem{hng1} H. Ghosh, Phys. Rev. B {\bf 60}, 3538 (1999).
\bibitem{hng2} H. Ghosh, Phys. Rev. B {\bf 63}, 226502 (2001).
\bibitem{hng3} H. Ghosh, Phys. Rev. B {\bf 59}, 3357 (1999).
\bibitem{Mazin}I. I. Mazin, D. J. Singh, M. D. Johannes and M. H. Du, Phys. Rev. Lett.{\bf 101}, 057003 (2008).
\bibitem{Barzykin}V. Barzykin and L. P. Gorkov, JETP Lett. {\bf 88}, 131 (2008).
\bibitem{Chubukov}A. V. Chubukov, D. Efremov and I. Eremin, Phys. Rev. B {\bf 78}, 134512 (2008).
\bibitem{Kontani}H. Kontani and S. Onari, Phys. Rev. Lett. {\bf 104}, 157001 (2010).
\bibitem{Yanagi}Y. Yanagi, Y. Yamakawa and Y. Ono, Phys. Rev. B {\bf 81}, 054518 (2010).
\bibitem{chris}A. D. Christianson, E. A. Goremychkin, R. Osborn, S. Rosenkranz, M. D. Lumsden, 
C. D. Malliakas, I. S. Todorov, H. Claus, D. Y. Chung, M. G. Kanatzidis, R. I. Bewley and 
T. Guidi, Nature {\bf 456}, 930 (2008).
\bibitem{Evtushinsky} D. V. Evtushinsky, D. S. Inosov, V. B. Zabolotnyy, 
M. S. Viazovska, R. Khasanov, A. Amato, H. -H. Klauss, H. Luetkens, Ch. Niedermayer, 
G. L. Sun, V. Hinkov, C. T. Lin, A. Varykhalov, A. Koitzsch, M. Knupfer, 
B. Bchner, A. A. Kordyuk and S.V.Borisenko, New J. Phys. {\bf 11}, 055069 (2009).
\bibitem{Bud’ko}S. L. Bud’ko, M. Sturza, D. Y. Chung, M. G. Kanatzidis and 
P. C. Canfield, Phys. Rev. B {\bf 79}, 220516 (2009).
\bibitem{Kim}J. S. Kim, G. R. Stewart, S. Kasahara, T. Shibauchi, T. Terashima and 
Y. Matsuda, J. Phys.: Condens. Matter {\bf 23}, 222201 (2011).
\bibitem{two-band-prb} O. V. Dolgov, I. Mazin, D. Parker and A. Golubov, Phys. Rev. B {\bf 79}, 060502(R) (2009).

% %\bibitem{Hammerath} Franziska Hammerath, Magnetism and Superconductivity in Iron-based Superconductors as Probed by Nuclear Magnetic Resonance, 2012.
\bibitem{Tao}T. Li, J. Phys.: Condens. Matter {\bf 20}, 425203 (2008).
\bibitem{Cao} C. Cao, P. J. Hirschfeld and H. -P. Cheng, Phys. Rev. B {\bf 77}, 220506(R) (2008).
\bibitem{Kuroki}K. Kuroki, S. Onari, R. Arita, H. Usui, Y. Tanaka, 
H. Kontani and H. Aoki, New J. Phys. {\bf 11}, 025017 (2009).
\bibitem{Raghu}S. Raghu, X. -L. Qi, C. -X. Liu, D. J. Scalapino and 
S. -C. Zhang, Phys. Rev. B, {\bf 77}, 220503(R) (2008).
\bibitem{Ghosh}H. Ghosh and H. Purwar, Europhys. Lett. {\bf 98}, 57012 (2012).
\bibitem{SSPS13}H. Ghosh, S. Sen, H. Purwar, AIP Conf. Proc. {\bf 1591}, 1621 (2014) .
\bibitem{natureFS}M. Sunagawa, T. Ishiga, K. Tsubota, T. Jabuchi, J. Sonoyama, K. Iba, 
K. Kudo, M. Nohara, K. Ono, H. Kumigashira, T. Matsushita, M. Arita, K. Shimada, H. Namatame, 
M. Taniguchi, T. Wakita, Y. Muraoka and T. Yokoya, Sci. Rep. {\bf 4}, 4381 (2014).
\bibitem{FS} Our first principle simulation on evaluation of FS for various 122 systems 
also confirms quasi-2d nesting, to be published elsewhere.
\bibitem{PRB2014}D. V. Evtushinsky, V. B. Zabolotnyy, T. K. Kim, A. A. Kordyuk, A. N. Yaresko, 
J. Maletz, S. Aswartham, S. Wurmehl, A. V. Boris, D. L. Sun, C. T. Lin, B. Shen, 
H. H. Wen, A. Varykhalov, R. Follath, B. Buchner and S.V.Borisenko, Phys. Rev. B, {\bf 89}, 064514 (2014).
\bibitem{Podolosky} D. Podolsky, H. -Y. Kee and Y. B. Kim, Europhys. Lett. {\bf 88}, 17004 (2009).
\bibitem{AIP}H. Ghosh and H. Purwar, AIP Conf. Proc. {\bf 328}, 1461 (2012).
\bibitem{Ding}H. Ding, P. Richard, K. Nakayama, T. Sugawara, T. Arakane, Y. Sekiba, 
A. Takayama, S. Souma, T. Sato, T. Takahashi, Z. Wang, X. Dai, Z. Fang, G. F. Chen, 
J. L. Luo and N. L. Wang, Europhys Lett. {\bf 83}, 47001 (2008).
\bibitem{Paglione}J. Paglione and R. L. Greene, Nature Phys. {\bf 6}, 645 (2010).
\bibitem{Gofryk}K. Gofryk, A. S. Sefat, E. D. Bauer, M. A. McGuire, B. C. Sales, 
D. Mandrus, J. D. Thompson and F. Ronning, New J. Phys. {\bf 12}, 023006 (2010).
\bibitem{Fang}C. Fang, H. Yao, W. -F. Tsai, J. P. Hu and S. A. Kivelson, arXiv: 0804.3843v1 (2008).
\bibitem{Davis}J. C. Davis and P. J. Hirschfeld, Nature Phys. {\bf 10}, 184 (2014).
\bibitem{Fernandes}R. M. Fernandes, A. V. Chubukov and J. Schmalian, Nature Phys. {\bf 10}, 97 (2014).
\bibitem{Nakajima}M. Nakajima, S. Ishida, Y. Tomioka, K. Kihou, C. H. Lee, 
A. Iyo, T. Ito, T. Kakeshita, H. Eisaki and S. Uchida, Phys. Rev. Lett. {\bf 109}, 217003 (2012).
\bibitem{Jiang}S. Jiang, H. S. Jeevan, J. Dong and P. Gegenwart, Phys. Rev. Lett. {\bf 110}, 067001 (2013).
\bibitem{Chu}J. -H. Chu, H. -H. Kuo, J. G. Analytis and I. R. Fisher, Science {\bf 337}, 710 (2012).
\bibitem{Ming}M. Yi, D. Lu, J. -H. Chu, J. G. Analytis, A. P. Sorini, A. F. Kemper, 
B. Moritz, S. -K. Mo, R. G. Moore, M. Hashimoto, W. -S. Lee, Z. Hussain, T. P. Devereaux, 
I. R. Fisher and Z. -X. Shen, Proc. Natl. Acad. Sci. U.S.A. {\bf 108}, 6878 (2011).



\end{thebibliography}
\end{document}